\documentclass[epj,nopacs]{svjour}

\usepackage{amssymb}
\usepackage{epsfig}
\usepackage{eucal}
\usepackage{amsmath}

\newcommand{\beq}{\begin{equation}} 
\newcommand{\eeq}{\end{equation}}   
\newcommand{\bea}{\begin{eqnarray}} 
\newcommand{\eea}{\end{eqnarray}}

\headnote{
\hfill \parbox{4cm}{\tt \normalsize IFIC/05-38 \\ TTP05-28 \\}}
 
\title{
 Electron-positron annihilation into three pions 
 and the radiative return.
\thanks{Work 
supported in part by BMBF under grant number 05HT4VKA/3,
EC 5th Framework Programme under contract HPRN-CT-2002-00311 
(EURIDICE network), TARI project RII3-CT-2004-506078,
Polish State Committee for Scientific Research (KBN)
under contract 1 P03B 003 28,
Ministerio de Educaci\'on y Ciencia under grant FPA2004-00996 (PARSIFAL),
and Generalitat Valenciana (GV05-015, GV04B-594 and GRUPOS03/013).}}

\author{Henryk Czy\.z\inst{1}
\thanks{\email{czyz@us.edu.pl}} 
\and
Agnieszka Grzeli{\'n}ska\inst{2}
\thanks{\email{grzel@joy.phys.us.edu.pl}}
\and
Johann H. K\"uhn\inst{2}
\thanks{\email{johann.kuehn@uni-karlsruhe.de}}
\and
Germ\'an Rodrigo\inst{3}
\thanks{\email{german.rodrigo@ific.uv.es}}
}

\institute{
Institute of Physics, University of Silesia, PL-40007 Katowice, Poland. 
\and
Institut f\"ur Theoretische Teilchenphysik,
Universit\"at Karlsruhe, D-76128 Karlsruhe, Germany.
\and
Instituto de F\'{\i}sica Corpuscular, CSIC-Universitat de Val\`encia,
Apartado de Correos 22085, E-46071 Valencia, Spain.
}

\date{May 2, 2006}

\abstract{
The Monte Carlo event generator PHOKHARA, which simulates hadron 
 and muon production
 at electron-positron colliders through radiative return, has been extended
 to final states with three pions. A model for the form factor based
 on generalized vector dominance has been employed, which is consistent
 with presently available experimental observations.
}

\begin{document}
\authorrunning{H. Czy\.z et al. }
\titlerunning{Electron-positron annihilation into three pions ...
   }

\maketitle

\section{Introduction}

Measurements of form factors and cross sections for electron positron
 annihilation in the low energy region provide important information
 on hadron dynamics. At the same time, they are necessary ingredients
in dispersion relations, which are used to predict hadronic contributions
to the momentum dependent electromagnetic coupling and the anomalous
 magnetic moment of the muon. The measurements are traditionally performed
 by tuning the center of mass energy of an electron positron collider
 to the point of interest. As an alternative, it has been advocated 
 to use the method of the radiative return \cite{Zerwas,Binner:1999bt} at
 high luminosity $\phi$- and $B$-meson factories. In this second case
 the collider energy remains fixed, while $Q^2$, the invariant mass
 of the hadronic system, can be varied by considering events, where
 one or several photons have been radiated. For a detailed and precise
 analysis initial and final state radiation must be included
 and a proper description of the various exclusive final states through
 appropriate form factors is required. All these ingredients are contained
 in the most recent version of the Monte Carlo event generator
 PHOKHARA4.0 \cite{Nowak,Czyz:PH04},
 which is based in particular on the virtual corrections described in
 \cite{Rodrigo:2001jr,Kuhn:2002xg} and
 which at present simulates production of
 $\mu^+\mu^-$, $\pi^+\pi^-$, four pions 
($2\pi^+2\pi^-$ and $\pi^+\pi^-2\pi^0$), $p\bar p$, and $n \bar n$
 \cite{Nowak,Czyz:PH04,Rodrigo:2001jr,Kuhn:2002xg,Rodrigo:2001kf,Czyz:2002np,Czyz:PH03}.
 In the present work the production of three pions is considered
 on the basis of form factors, which include already available information
 on the production cross section and on differential distributions
 in two-pion subsystems. The model implements three-pion production
 through $\omega$, $\phi$ and their radial excitations and the subsequent
decay of these resonances into $\rho\pi$, $\rho'\pi$ and $\rho''\pi$.
 A small isospin-violating component $\gamma^*\to\omega(\to\pi^+\pi^-)\pi^0$
 is added, which is needed to properly describe the data. 
 The total cross section and the distributions are well reproduced
 within this model. It is furthermore demonstrated that the couplings
 introduced and adopted for this purpose lead to a satisfactory
 description of $\Gamma(\pi^0\to\gamma\gamma)$, of the slope parameter
 of the $\pi^0\to\gamma\gamma^*$ amplitude and of the radiative 
 vector meson decays $\rho\to \pi^0\gamma$, $\phi\to \pi^0\gamma$,
 but is in conflict with  $\omega\to \pi^0\gamma$.


\section{A phenomenological description of three-pion production}

 The amplitude for three-pion production through the electromagnetic current
 is restricted by current conservation and negative parity
 to the form

\bea
J_{\nu}^{\mathrm{em},3\pi} &=& 
 \langle\pi^+(q_+) \ \pi^-(q_-) \ \pi^0 (q_0)|J_{\nu}^{\mathrm{em}}
|0\rangle\nonumber \\
 &=& \epsilon_{\nu \alpha \beta \gamma} q_{+}^{\alpha} 
q_{-}^{\beta} q_{0}^{\gamma} \ F_{3\pi}(q_{+},q_{-},q_{0}) \ .
\label{3pi_p}
\eea

\begin{figure}[ht]
\begin{center}
\includegraphics[width=8.5cm,height=4.5cm]{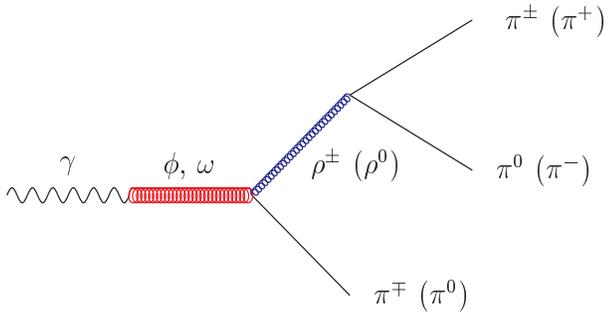}
\caption{Diagrams contributing to the 3--pion current: $I=0$ component.}
\label{3pi_diag}
\end{center}
\end{figure}

G-parity dictates dominance of the isospin-zero component of 
the electromagnetic current , which will be discussed in a first step.
 The small isospin-one admixture will be discussed subsequently.
 The form factor $F_{3\pi}^{I=0}$ is constructed under the
 assumption that the virtual photon couples to the $\omega$-
 and $\phi$-meson, whose subsequent transition to three pions
 is dominated by  the $\rho(\to 2\pi)\pi$ chain (Fig.\ref{3pi_diag})
 \cite{Gell-Mann}.
 Taking into account radial excitations of $\omega$, $\phi$, $\rho$
 one arrives at the form factor

\bea
\kern-15pt F_{3\pi}^{I=0}&&(q_{+},q_{-},q_{0}) =  \nonumber \\
 &&\sum_{i,j}a_{ij} \cdot BW_{V_i}(Q^2)
   \cdot H_{\rho_j}\left(Q_+^2,Q^2_-,Q_0^2\right)
 \ ,
\label{3pi_pb}
\eea
where $V_i$ stands for either $\omega$- or $\phi$-resonances,
 and $\rho_j$ represents contributions from $\rho$-mesons.
 From the PDG \cite{PDG04} it is clear that all
 $\omega$- and $\phi$-resonances couple to the ground state
 of the $\rho$-meson, whereas
 there is no indication about couplings to the higher radial excitations 
 ($\rho',\rho'',\cdots$). This missing piece of information
 can, however, be obtained to large extent from the 
  known $e^+e^-\to \pi^+\pi^-\pi^0$ cross section, as shown below.

 For the function $H_\rho$ we shall adopt the ansatz 

\bea
H_{\rho}(Q_{+}^{2},Q_{-}^{2},Q_{0}^{2}) = BW_{\rho}(Q_{0}^{2}) 
+ BW_{\rho}(Q_{+}^{2}) + BW_{\rho}(Q_{-}^{2}) \ , \nonumber \\ 
\phantom{}\kern-15pt
\eea

\noindent
with

\bea
Q_{0}^{2} = (q_{+}+q_{-})^2, \ 
Q_{\pm}^{2} = (q_{\mp}+q_{0})^2, \ 
\eea
and the Breit-Wigner form factors are
\bea
BW_{V}(Q^2) &=& \biggl[ \frac{Q^2}{m_V^2} - 1 + i
\frac{\Gamma_V}{m_V} \biggr]^{-1} \ , \nonumber \\
BW_{\rho}(Q^{2}_{i}) &=& \biggl[ \frac{Q^{2}_{i}}{m_{\rho}^2} - 1
+ i \frac{\sqrt{Q^2_{i}}\Gamma_{\rho}(Q^{2}_{i},m_j,m_k) }{m_{\rho}^2} \biggr]^{-1}
 ,
\label{BW}
\eea
where $Q_i^2 = (q_j + q_k)^2$, and $m_{j} = m_{\pi^j}$, with
 $i,j,k = 0,\pm$. 
 We use
propagators with constant widths for $\omega$'s  and $\phi$, and
energy dependent widths for $\rho$- resonances as predicted by P-wave
 $\rho\to\pi\pi $ decays:

\bea
\Gamma_{\rho}(Q^2_{i},m_j,m_k) = \Gamma_{\rho} \frac{m_{\rho}^2}
{Q^2_{i}}\biggl[
\frac{Q^2_{i} - (m_j+m_k)^2 }
     {m_{\rho}^2 - (m_j+m_k)^2 } \biggr]^{3/2} \ . 
\eea
The couplings $a_{ij}$ are taken as real constants and 
we assume that the isospin symmetry is violated in this component only by the
 $\pi^0-\pi^\pm$ mass difference. 

\begin{figure}[ht]
\begin{center}
\epsfig{file=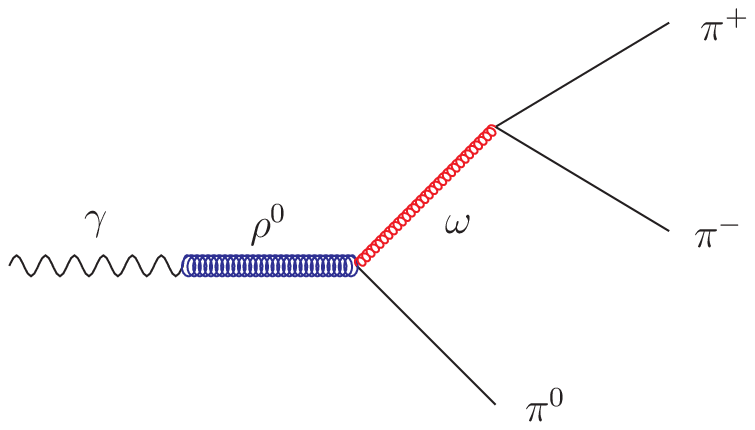,width=8.cm} 
\caption{Diagram contributing to the $I=1$ component
 of the three-pion current.}
\label{3pi_diag1}
\end{center}
\end{figure}
 The small isospin violating amplitude is mediated by the $I=1$
 component of the electromagnetic current and is based on
the $4\pi$ current of Refs. \cite{Czyz:2000wh,Decker}
 i.e. we take the $\rho-\gamma$ and $\rho\pi\omega$ couplings
 from the $4\pi$ current and replace the $\omega\to 3\pi$ transition
 by the isospin violating $\omega\to 2 \pi$ decay
 as shown  in Fig.\ref{3pi_diag1}.
This leads to the following ansatz
\bea
F_{3\pi}^{I=1}(q_{+},q_{-},&&q_{0}) = G_{\omega}
 \cdot BW_{\omega}(Q^2)/\tilde m^2_{\omega}\nonumber \\
&&\kern-15pt\left[BW_{\rho}(Q_0^2)/\tilde m^2_\rho+\sigma BW_{\rho''}(Q_0^2)/
\tilde m^2_{\rho''}\right] \ \ ,
\label{i0}
\eea
\noindent
where
\bea
G_{\omega}= \frac{1.55}{\sqrt{2}}\ 12.924\ {\rm GeV}^{-1} \ 0.266 \ m_\rho^2 
 \ g_{\omega\pi\pi}
\label{i0p}
\eea
\noindent
and $\tilde m_\rho = 0.77609$ GeV, $\tilde\Gamma_\rho = 0.14446$ GeV, 
$\tilde m_{\rho''} = 1.7$ GeV, $\tilde\Gamma_{\rho''} = 0.26$ GeV,
 $\sigma=-0.1$, where the parameters are taken directly 
 from \cite{Decker}.
 At the present level of experimental accuracy it is not clear
 whether the $\rho''$ term is necessary for the description of
  the 3$\pi$ current (see 
 below). However, as it is a prediction coming from the 4$\pi$ current
 we consider its contribution also here. 
\begin{figure*}[ht] 
\begin{center}
\epsfig{file=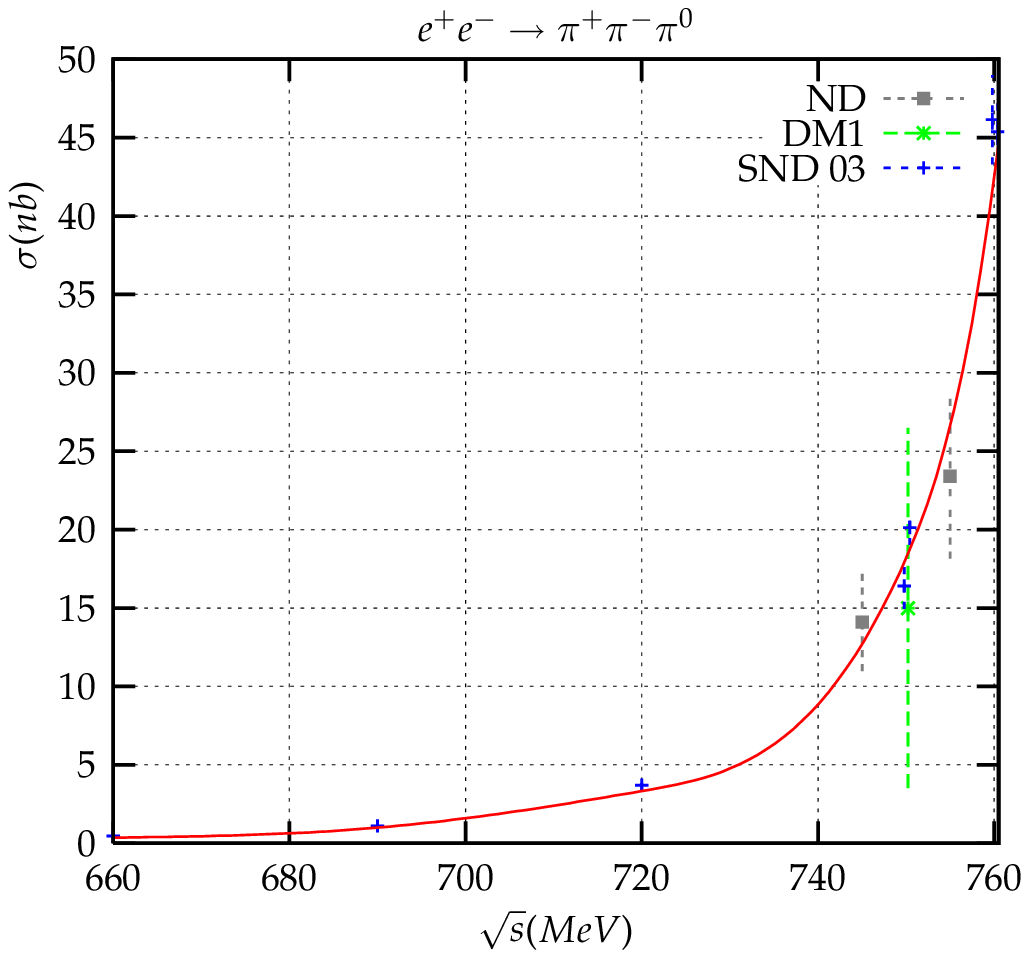,width=8.25cm}
\epsfig{file=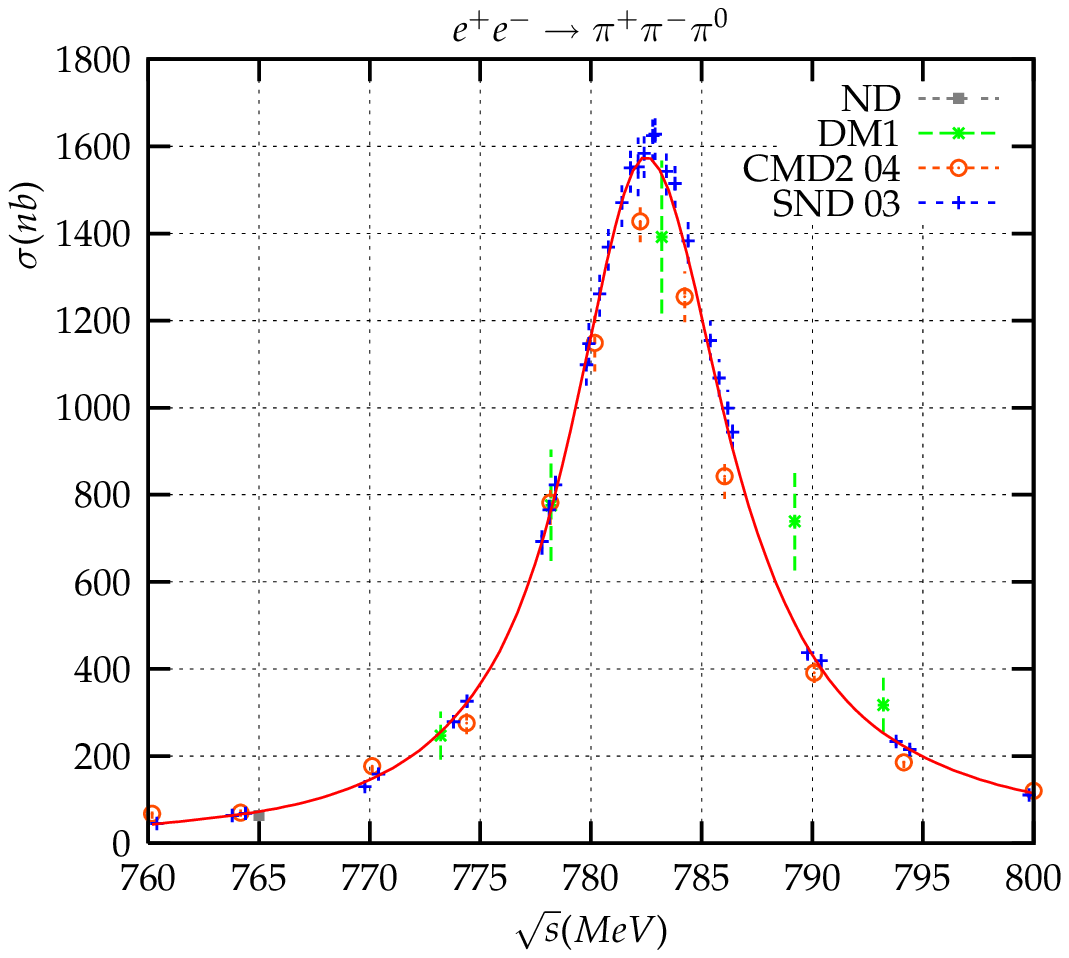,width=8.5cm}
\caption{$e^+e^-\to \pi^+\pi^-\pi^0$ cross section obtained with 
 fitted parameters (solid line, see text for details) vs. experimental data.}
\label{date:1}
\end{center}
\end{figure*}
\begin{figure*}[ht] 
\begin{center}
\epsfig{file=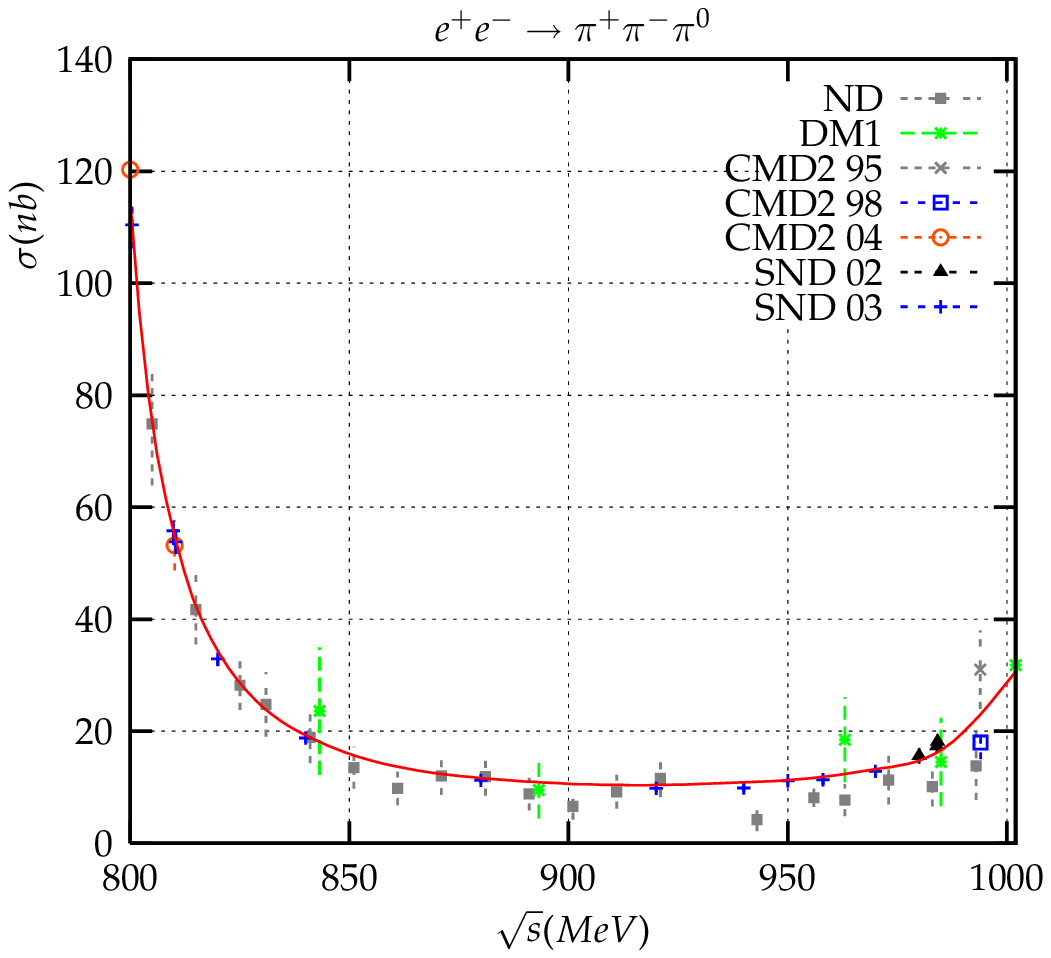,width=8.5cm}
\epsfig{file=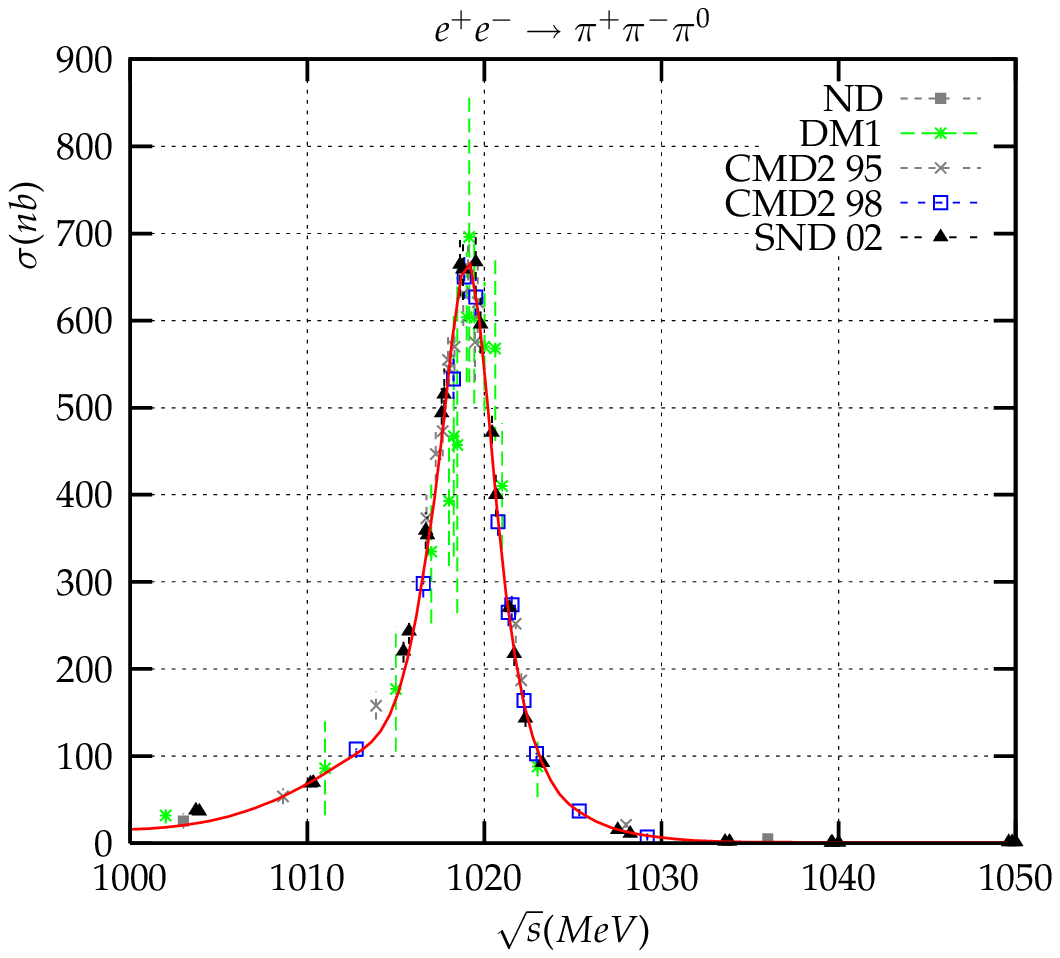,width=8.5cm}
\caption{$e^+e^-\to \pi^+\pi^-\pi^0$ cross section obtained with 
 fitted parameters (solid line, see text for details) vs. experimental data.}
\label{date:2}
\end{center}
\end{figure*}

\begin{figure}[ht] 
\begin{center}
\epsfig{file=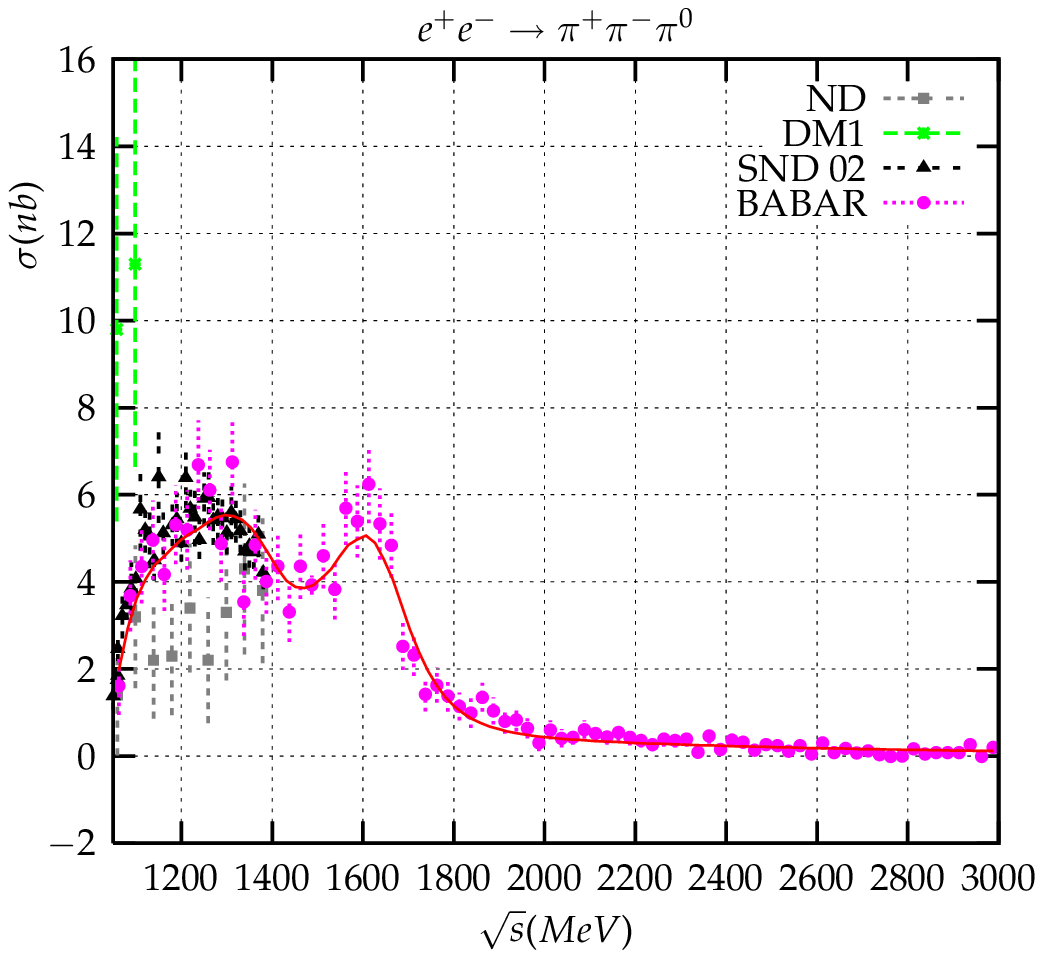,width=8.5cm}
\caption{$e^+e^-\to \pi^+\pi^-\pi^0$ cross section obtained with 
 fitted parameters (solid line, see text for details) vs. experimental data. }
 \label{date:3}
\end{center}
\end{figure}

The coupling $g_{\omega\pi\pi}$ can be extracted from the 
 decay rate $\Gamma(\omega\to\pi\pi)$ 
  (see Eq.(\ref{gamrhopipi})).
Using the world average value from the PDG \cite{PDG04} 
 one gets $g_{\omega\pi\pi}= 0.185(15)$. 
 The total form factor is of course given by the coherent sum
\bea
 \kern-15pt F_{3\pi}(q_{+},q_{-},q_{0}) =  F_{3\pi}^{I=0}(q_{+},q_{-},q_{0})
  + F_{3\pi}^{I=1}(q_{+},q_{-},q_{0}) \ .
\label{3pi_pa}
\eea

Data on $\sigma(e^+e^-\to\pi^+\pi^-\pi^0)$, from 
energy scan experiments
\cite{CMD2,SND,Dol:91,DM1,Antonelli:92} 
consist of 217 data points, covering the energy range from $~660$ MeV 
 to $~2400$ MeV. Moreover, by using the radiative return method,
  BaBar \cite{babar3pi} has obtained additional 78  data points, 
  covering
 the energy from 1.06 GeV up to almost 3 GeV.
Data from the DM2 collaboration 
\cite{Antonelli:92} being inconsistent
 with the more accurate BaBar data, were not used in the fit.

\begin{table}[hb]
\begin{center}
\caption{Values of the couplings masses and widths obtained in the fit;
 couplings $A-F$ in ${\rm GeV}^{-3}$ masses and widths
 in ${\rm MeV}$ (see text for details).}
\label{tab:cons_F_c}
\begin{tabular}{l|l|l|l}\hline
$m_{\omega(782)}$  & 782.4(4)    & $A $ &   18.20(8)   \\ \hline
$\Gamma_{\omega(782)}$  & 8.69(7)& $B $ &  -0.87(5)   \\ \hline
$m_{\phi(1020)}$  & 1019.24(3)  & $C $ &  -0.77(5)  \\ \hline
$\Gamma_{\phi(1020)}$  & 4.14(5) & $D $ &  -1.12(4)   \\ \hline
$m_{\omega(1420)}$  &  1375(1)   & $E $ &  -0.72(10)  \\ \hline
$\Gamma_{\omega(1420)}$  & 250(5)& $F $ &  -0.59(4)  \\ \hline
$m_{\omega(1650)}$ &  1631(6)    & &   \\ \hline
$\Gamma_{\omega(1650)}$ & 245(13)& $\chi^2/\mathrm{d.o.f}$& 1.14  \\ \hline
\end{tabular}
\end{center}
\end{table}
\begin{figure}[ht] 
\begin{center}
\epsfig{file=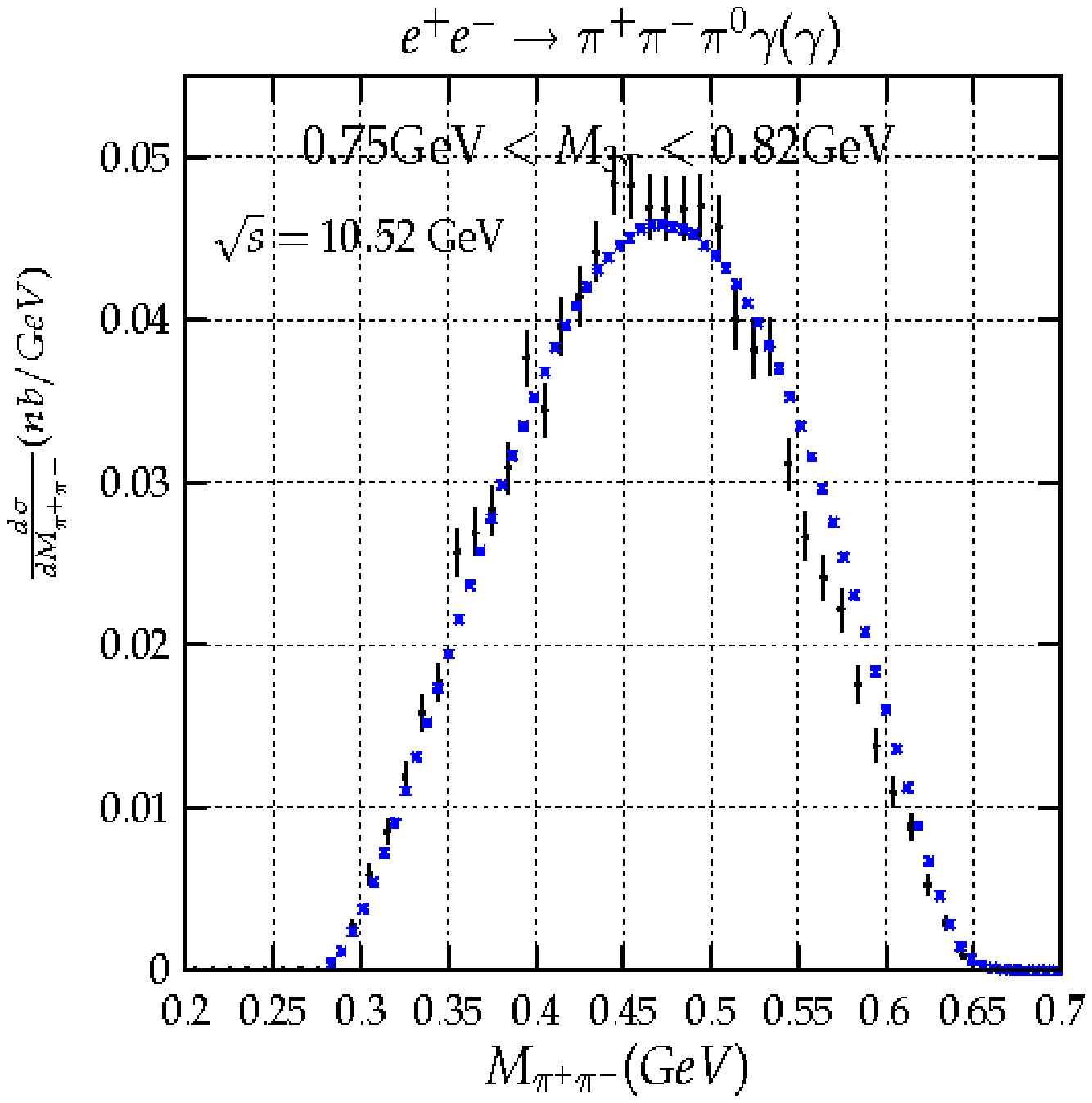,width=4.1cm}
\epsfig{file=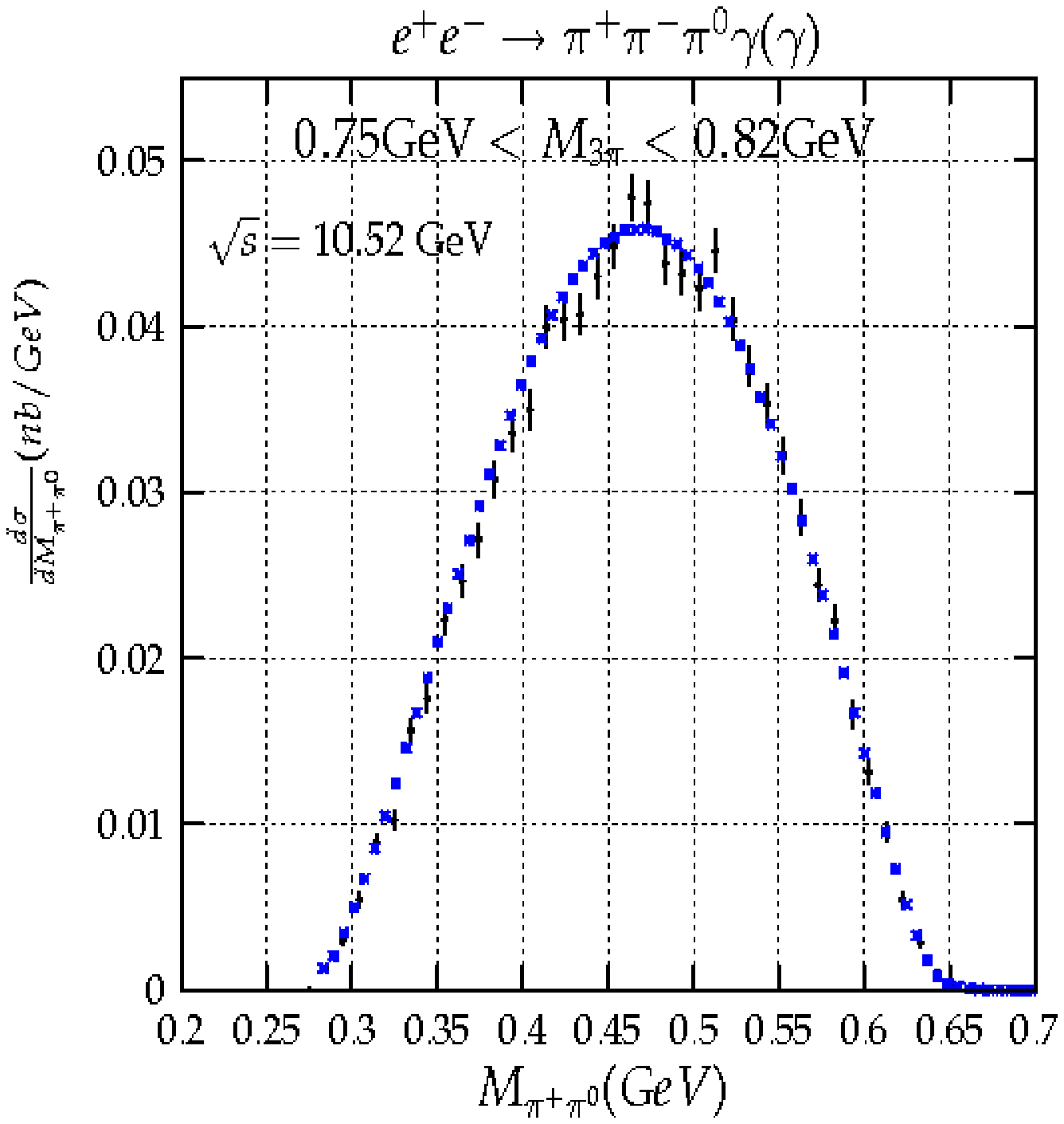,width=4.cm} \\
\epsfig{file=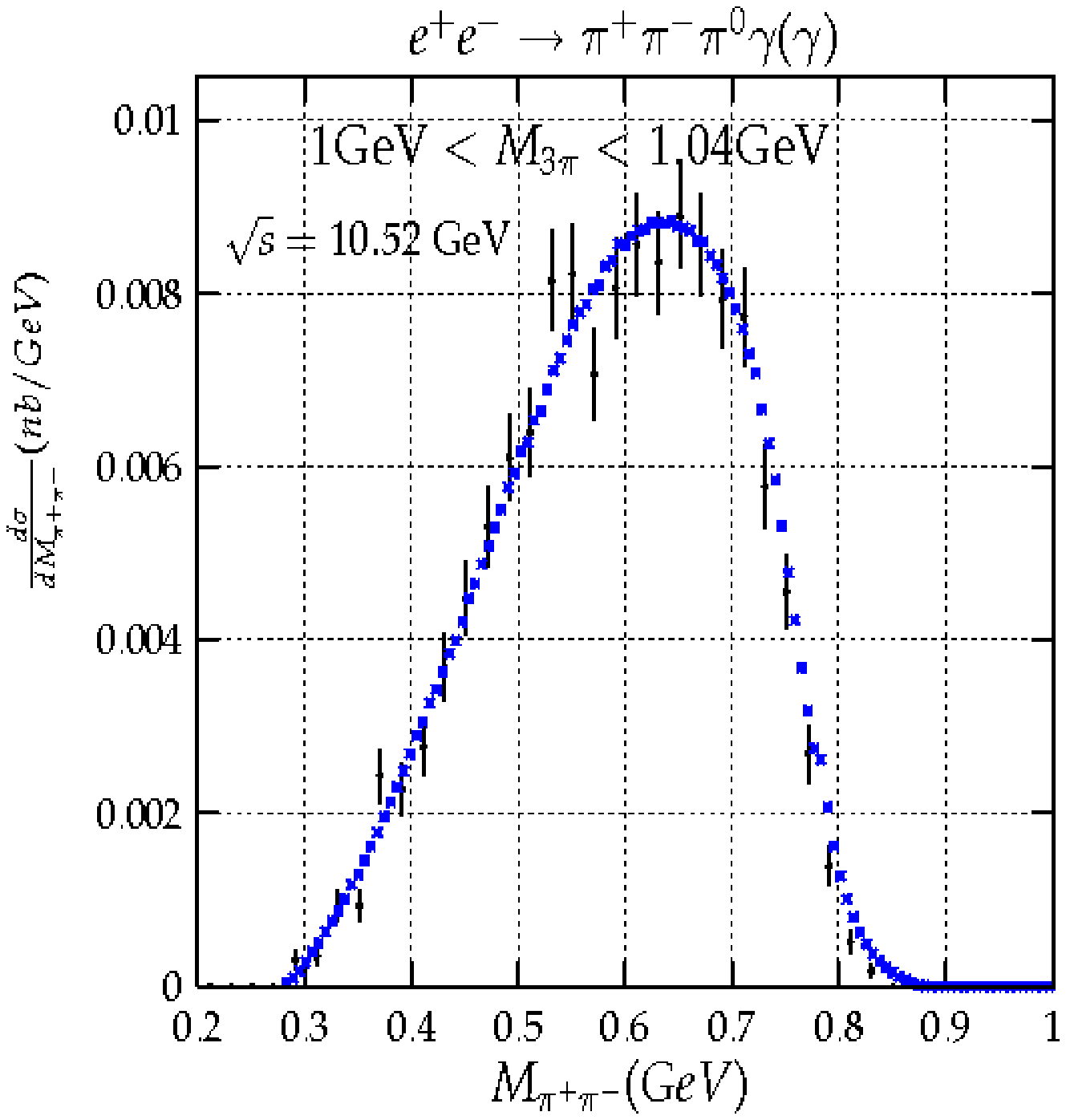,width=4.05cm}
\epsfig{file=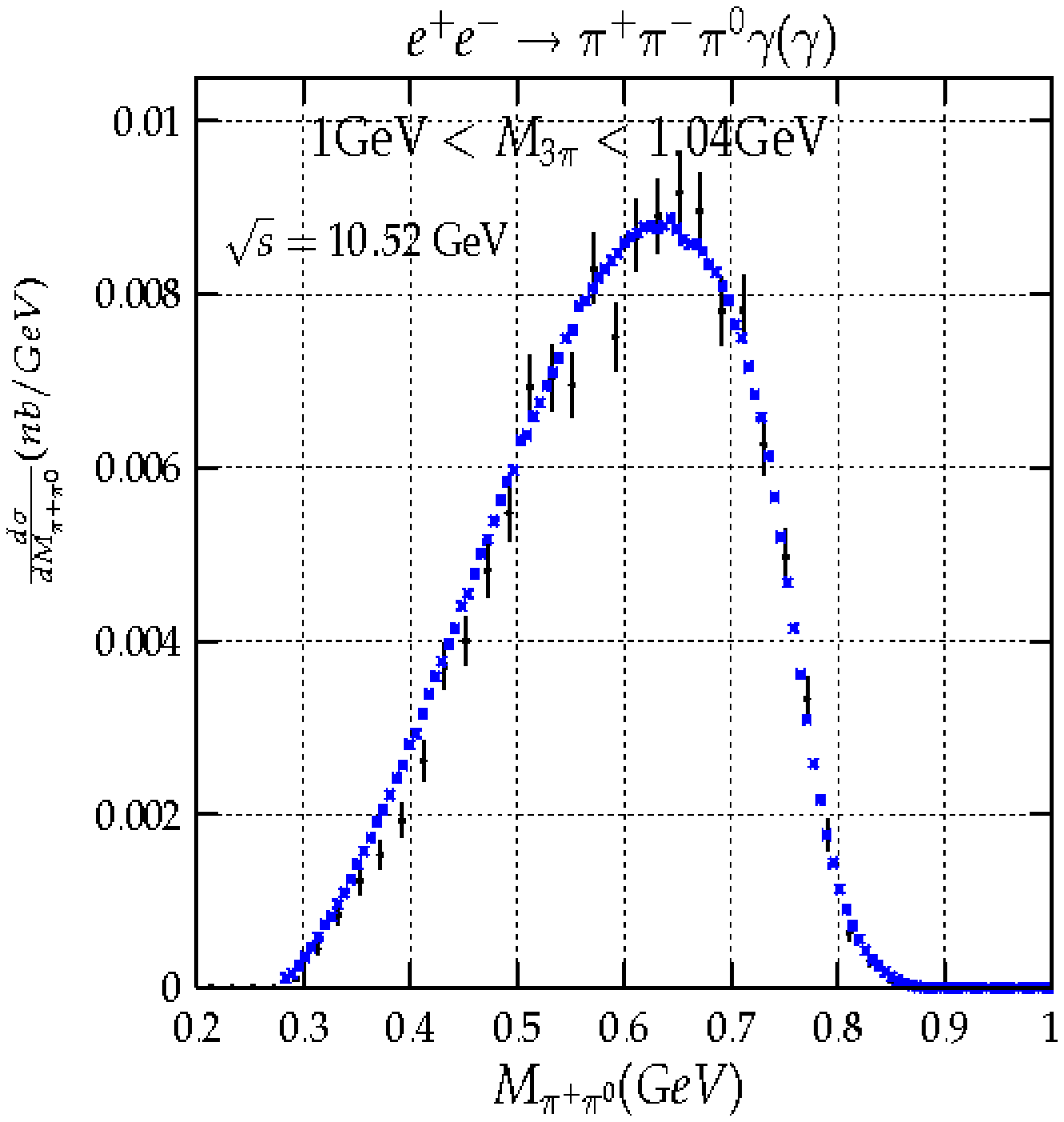,width=4.cm}\\
\epsfig{file=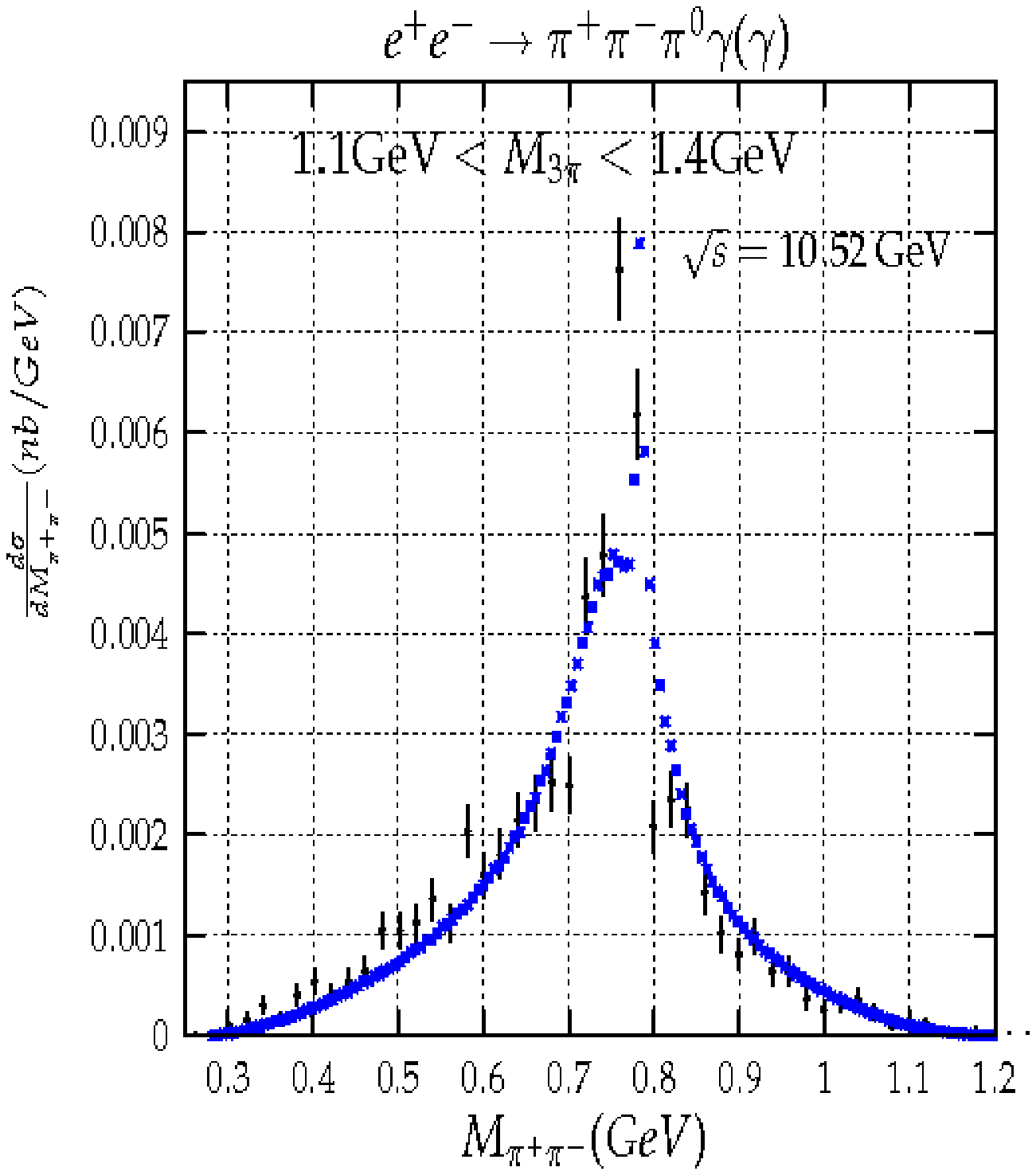,width=4.0cm}
\epsfig{file=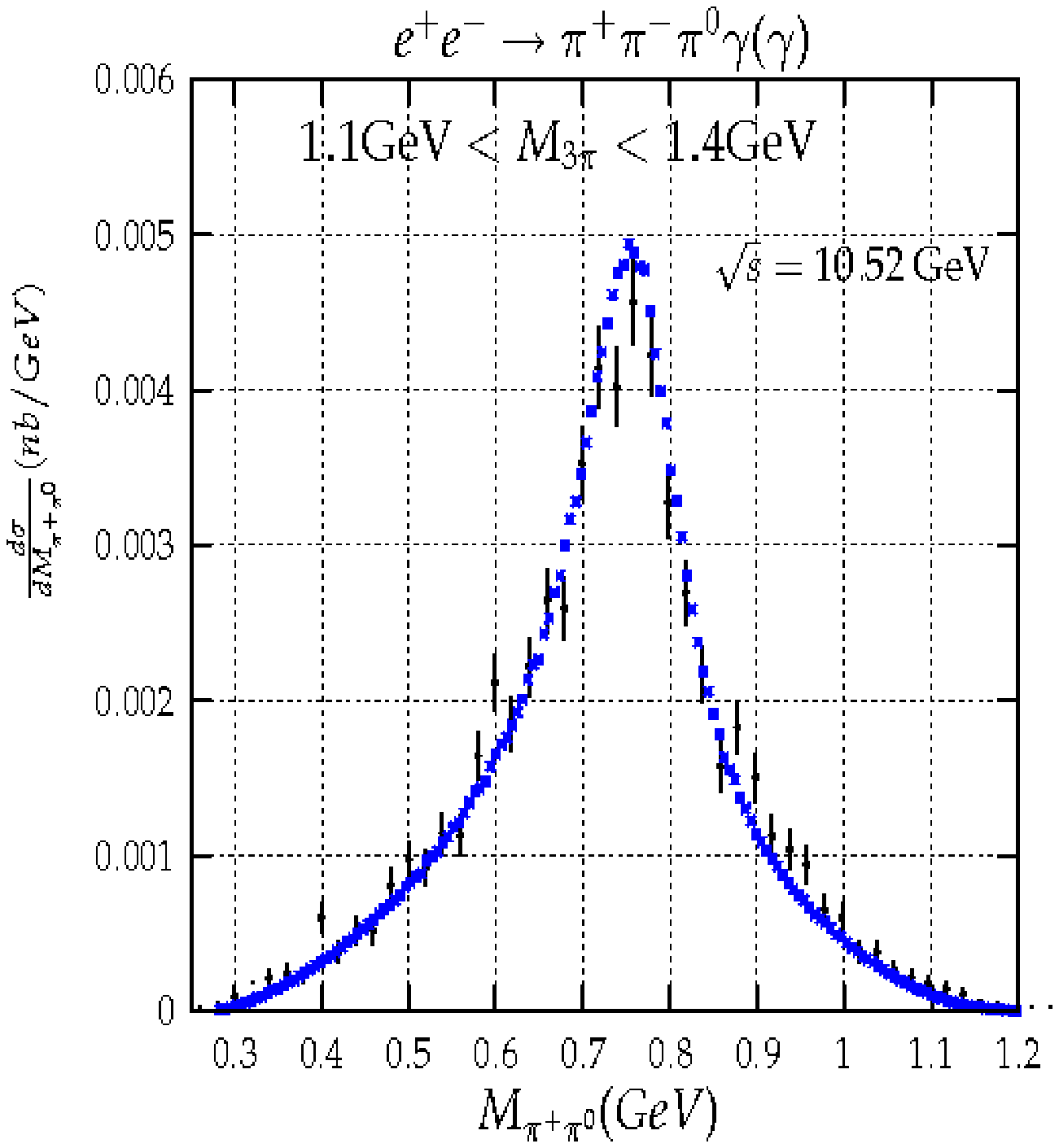,width=4.2cm}\\
\includegraphics[width=3.8cm,height=4.55cm]{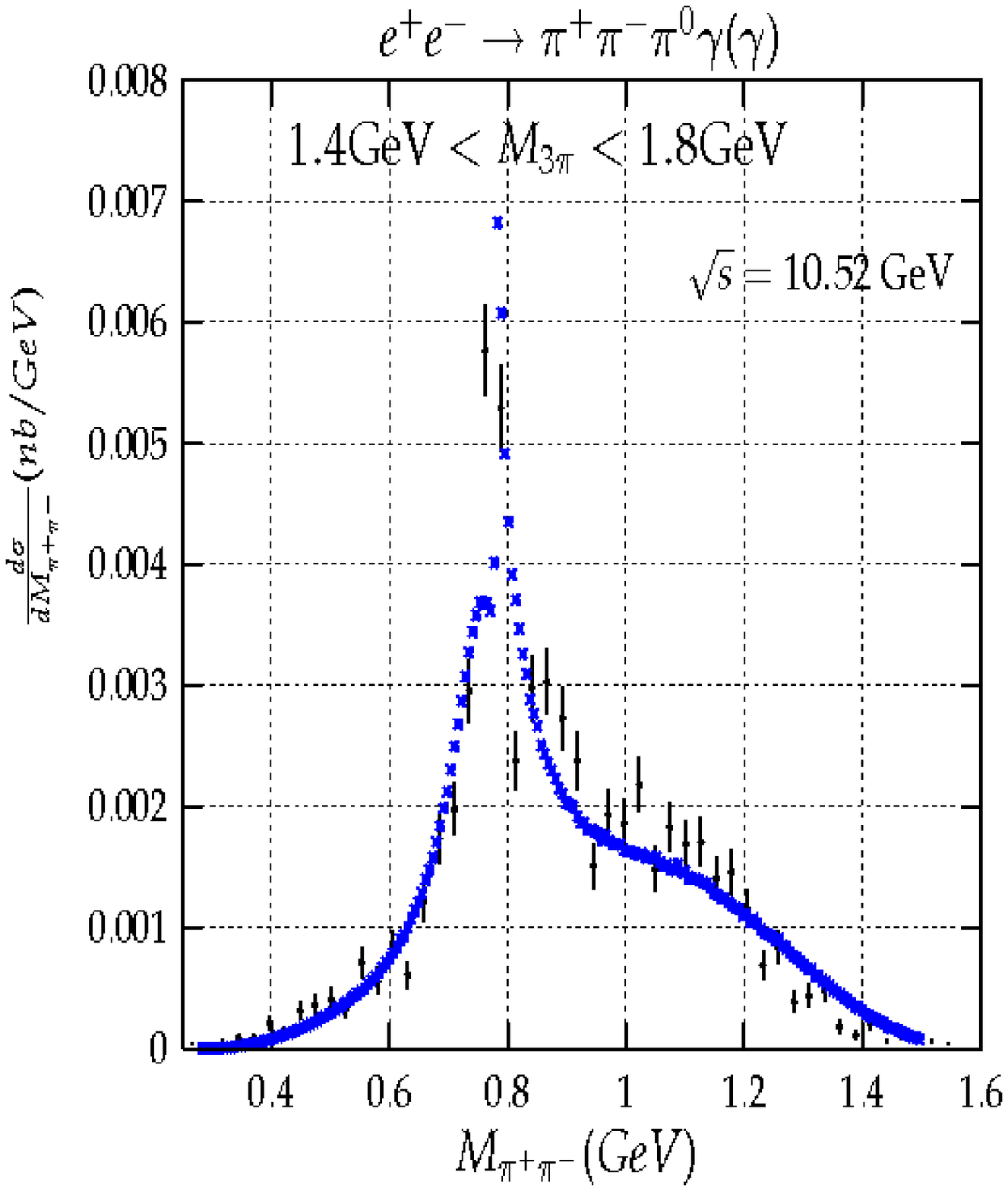}
\includegraphics[width=3.8cm,height=4.7cm]{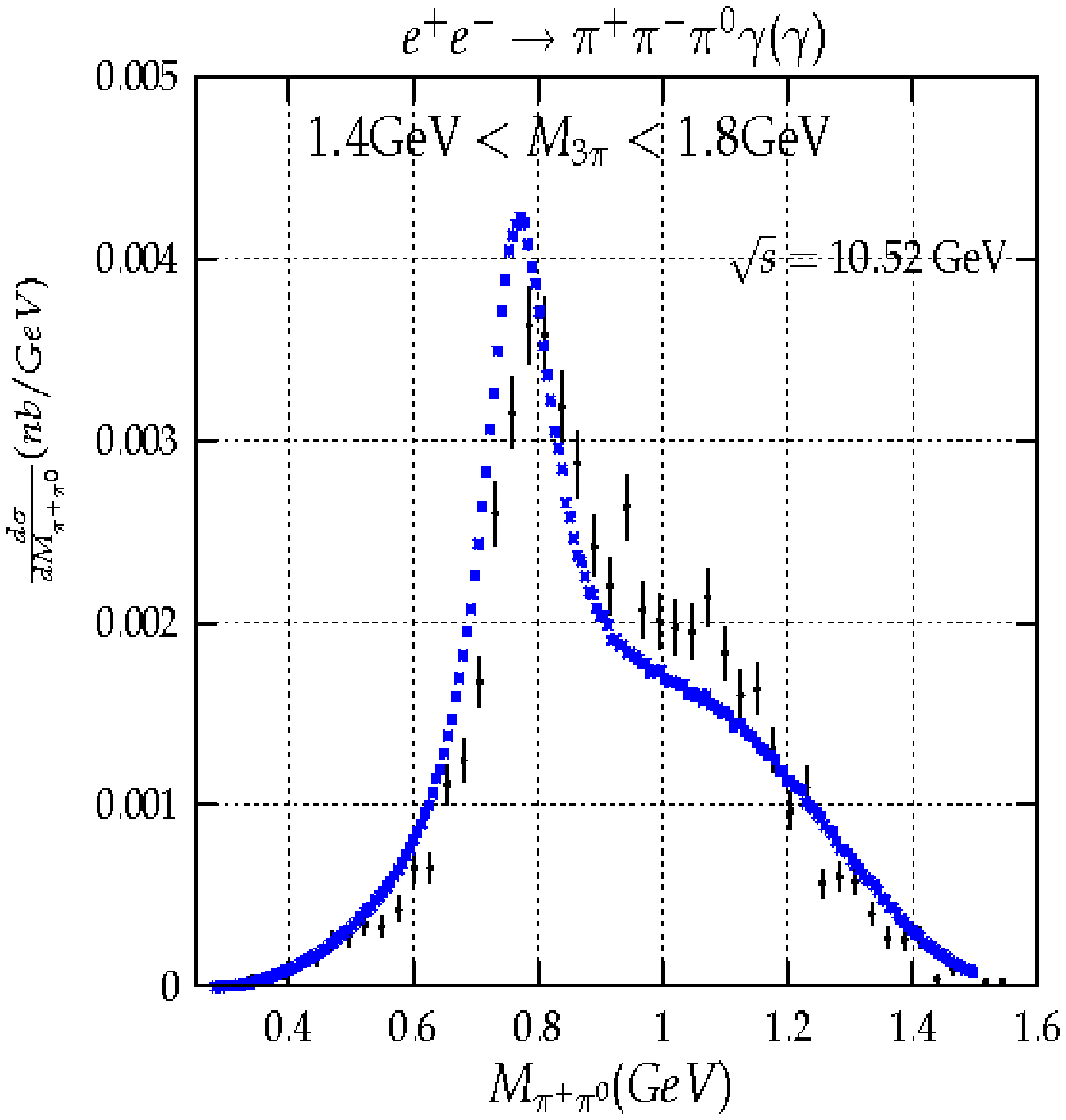}
\caption{Two pion invariant mass distributions for four different
 ranges of   $\pi^+\pi^-\pi^0$ invariant mass. 
  The BaBar data points, given as events/bin,
 are superimposed
 on plots obtained by PHOKHARA
 (see text for details). }
 \label{subdistr}
\end{center}
\end{figure}

 To fit the experimental data we include the following 
 resonances:
 $\omega(782)$, $\omega'\equiv \omega(1420)$, $\omega''\equiv\omega(1650)$,
 $\phi(1020)$, $\rho(770)$,
 $\rho'\equiv \rho(1450$) and $\rho''\equiv \rho(1700)$.
 The strategy was to minimize the number of
  contributions in Eq.(\ref{3pi_pb}) and arrive 
 at a good fit in the same time.
 Our best fit for
  the isospin zero component reads 
 \begin{align}
& F_{3\pi}^{I=0}(q_{+},q_{-},q_{0}) = 
H_{\rho(770)} (Q_{+}^{2},Q_{-}^{2},Q_{0}^{2})\nonumber \\ &
  \kern+50pt \cdot  \biggl[ 
 { A \cdot BW_{\omega(782)}}(Q^2)
+  B \cdot {  BW_{\phi(1020)}} (Q^2) \nonumber \\ &
\kern+50pt +  C \cdot  BW_{\omega(1420)} (Q^2)
+  D \cdot  BW_{\omega(1650)} (Q^2)
\biggr]  \nonumber \\ &
\kern+20pt 
 + E \cdot {  BW_{\phi(1020)}} (Q^2)\cdot   
 H_{\rho(1450)} (Q_{+}^{2},Q_{-}^{2},Q_{0}^{2})\nonumber \\ &
\kern+20pt + F \cdot BW_{\omega(1650)} (Q^2)
 \cdot H_{\rho(1700)} (Q_{+}^{2},Q_{-}^{2},Q_{0}^{2})
  \ ,
\label{3pi_fac}
\end{align}
with  the couplings and masses
given in Table \ref{tab:cons_F_c}. 
The errors are the
parabolic errors calculated by the MINUIT processor MINOS and correspond
to the change of $\Delta\chi^2=1$. 
The value
$\chi^2/\mathrm{d.o.f} =1.14$ is at the edge of the 95\% confidence 
interval. It is to large extent a result of the fact that the data 
 of different experiments are only marginally mutually consistent
 as evident from Figures \ref{date:1}--\ref{date:3}.
 Lack of published numerical information on differential distributions 
 does not allow for further refinements
of the model. In particular, indications for additional contributions
from higher radial excitations in the region of large $Q^2$ cannot be
substantiated by more detailed fits.
Access to distributions in invariant 
masses of pion pairs and to pion angular distributions would allow
to study these effects.
\begin{table}[ht]
\begin{center}
\caption{The masses  and widths of $\rho$ resonances used in the fits}
\label{rho}
\begin{tabular}{l|l|l|l} 
 & $\rho(770)$& $\rho(1450)$&$\rho(1700)$ \\ \hline
$m$ (GeV) & 0.77609& 1.465& 1.7\\ \hline
$\Gamma$ (GeV) & 0.14446& 0.31&0.235 \\ \hline
\end{tabular}
\end{center}
\end{table} 
For the moment we
 set the masses and widths of the $\rho$ mesons to the values  
collected in Table \ref{rho} and  assume equal masses and widths
 for the neutral
 and charged $\rho$ mesons.
Nevertheless qualitative 
comparisons of two--pion invariant mass distributions
can be performed by superimposing the experimental and the model
(data from Fig. 15 of \cite{babar3pi} and differential cross sections
 generated with 
 PHOKHARA 5.0)
 distributions (see Fig.\ref{subdistr}).
The $I=0$ component alone predicts identical distributions in 
 $M_{\pi^+\pi^-}$ and $M_{\pi^\pm\pi^0}$. The small $I=1$ component
 is concentrated in the spike at  $M_{\pi^+\pi^-}=m_\omega$.
 This channel starts to contribute for $Q^2$ values above 1~GeV only.
 The model seems to describe the distributions reasonably well.
 However the following deviations are observed:
 In the lowest region (0.75~GeV  $< M_{3\pi} <$ 0.82~GeV) the experimental
 results for the distributions in $M_{\pi^+\pi^-}$ and $M_{\pi^\pm\pi^0}$,
 respectively,
 seem to differ in the upper range, an indication of isospin-violation,
 that cannot be reproduced by our ansatz. In the large $Q^2$ range
 (1.4~GeV  $< M_{3\pi} <$ 1.8~GeV) an excess is observed in both charge
 modes for masses of the two-pion system between 1~GeV and 1.2~GeV.
 A similar excess is not observed in the pion form factor.  

%


\section{Meson couplings and  partial decay widths.}

 From the results of the fit we can evaluate the
 meson couplings separately, combining the following relations
\bea
A&=&2 g_{\omega\gamma}\cdot g_{\omega\pi\rho}\cdot g_{\rho\pi\pi} 
 \nonumber \\
B&=&2 g_{\phi\gamma}\cdot g_{\phi\pi\rho}\cdot g_{\rho\pi\pi} 
 \nonumber \\
C&=&2 g_{\omega'\gamma}\cdot g_{\omega'\pi\rho}\cdot g_{\rho\pi\pi}
 \nonumber \\
D&=&2 g_{\omega''\gamma}\cdot g_{\omega''\pi\rho}\cdot g_{\rho\pi\pi} 
 \nonumber \\
E&=&2 g_{\phi\gamma}\cdot g_{\phi\pi\rho'}\cdot g_{\rho'\pi\pi}
 \nonumber \\
F&=&2 g_{\omega''\gamma}\cdot g_{\omega''\pi\rho''}\cdot g_{\rho''\pi\pi} \ ,
 \label{couplings}
\eea
 with information about partial decay widths.
From the known decay widths of $\rho^0,\omega$ and $\phi$ to two pions
 one determines the couplings $g_{V\pi\pi}$ ($V = \rho^0,\omega,\phi$):
\bea
\Gamma_{V\to \pi^+\pi^-} = g_{V\pi\pi}^2 \frac{m_V}{48\pi}
 \left[1-\frac{4m_{\pi^+}^2}{m_V^2}\right]^{3/2} \ .
\label{gamrhopipi}
\eea
Similarly, the $g_{V\gamma}$ couplings can be found from the measured
values of  $\Gamma(V\to e^+e^-)$:
\bea
\Gamma_{V\to e^+e^-} = g_{V\gamma}^2 \frac{4\pi\alpha^2}{3m_V^3} \ .
\label{gamrhoee}
\eea
 For the $\omega'$ and $\omega''$ we do not know
 the partial decay widths and the couplings cannot be extracted 
 separately. The numerical values of the couplings 
obtained from Eq.(\ref{couplings})
 and partial decay widths \cite{PDG04} are collected in Table 
 \ref{tab:couplings}.

\begin{table}[hb]
\begin{center}
\caption{Values of the three-- and two--particle couplings;
\ \  $g_{V\pi\pi}$ is dimensionless, $g_{V\gamma}$ in ${\rm GeV}^{2}$,
 $g_{V\pi\rho}$ in ${\rm GeV}^{-5}$.}
\label{tab:couplings}
\begin{tabular}{l|l|l|l}\hline
$g_{\rho\pi\pi}$    &  5.997(32) & $g_{\rho\gamma}   $ &  0.1212(13) 
\\ \hline
$g_{\omega\pi\pi}$  &  0.185(15) & $g_{\omega\gamma} $ &  0.03591(37)
\\ \hline
$g_{\phi\pi\pi}$    &  0.0072(6) & $g_{\phi\gamma}   $ & 0.0777(7)
 \\ \hline
 $g_{\omega\pi\rho}$ &   42.3(5) &$g_{\omega'\gamma}\cdot g_{\omega'\pi\rho}$ & -0.064(8) \\ \hline
$g_{\phi\pi\rho}$ &  -0.93(5)
& $g_{\omega''\gamma}\cdot g_{\omega''\pi\rho}$ &  -0.093(3) \\ \hline
\end{tabular}
\end{center}
\end{table}

 Having extracted the couplings, we are able to predict many 
 physical quantities that have been measured
 already and check the model. 
 In particular we will investigate if various meson--photon interactions
 can be modeled, as in the $3\pi$ case, by three--meson couplings and 
 vector meson-photon mixings with the couplings and propagators as 
 introduced above 
(for a review of alternative models see \cite{BGP_DAPHNE,Harada}). 
 As one can recognize our model is an  extension to higher radial
 excitations of the model outlined in \cite{Gell-Mann}. 
 
\begin{figure}
\begin{center}
\includegraphics[width=4.cm,height=4cm]{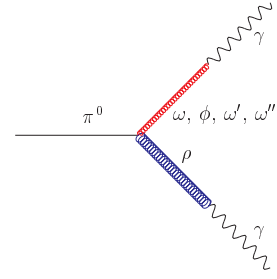}
\caption{Diagrams contributing to $\pi^0\to\gamma\gamma$ decay.}
\label{pi0gamgma}
\end{center}
\end{figure}

 Let us start with the decay width of $\pi^0\to\gamma\gamma$. 
 The model is based on the diagrams 
  shown in Fig.\ref{pi0gamgma}. As indicated by the fit,
 contributions from  $\rho'-\gamma$ and $\rho''-\gamma$ mixings are
 not required at the present level of precision.

\begin{table}[hb]
\begin{center}
\caption{ Mean life time for $\pi^0\to 2\gamma$ in seconds and decay rates 
 for $\rho,\omega,\phi \to \pi^0\gamma$ in MeV as
 obtained within our model compared to experimental
 results \cite{PDG04}.}
\label{direct}
\begin{tabular}{l|l|l} 
  & model & experiment \\ \hline
 $\tau(\pi^0\to\gamma\gamma)$ & 6.6(3)$\cdot 10^{-17}$ 
   &  8.3(6)$\cdot 10^{-17}$\\ \hline
 $\Gamma(\rho^0\to\pi^0\gamma)$ & 0.078(3)
   & 0.090(20) \\ \hline
 $\Gamma(\omega\to\pi^0\gamma)$
  & 1.31(4) 
  & $0.757^{+25}_{-22}$\\ \hline
 $\Gamma(\phi\to\pi^0\gamma)$ & 4.2(5)$\cdot 10^{-3}$
 & 5.2(4)$\cdot 10^{-3}$\\ \hline
\end{tabular}
\end{center}
\end{table} 
The partial decay width of the decay $\pi^0\to\gamma\gamma$ is thus given
 in our model by 
\bea
\Gamma_{\pi^0\to \gamma\gamma} &=&  \pi\alpha^2 m_{\pi^0}^3\
   g^2_{\rho\gamma}  \
 T^2\ ,
\label{gampi0gg}
\eea
 with 
\bea
 T=
  g_{\omega\pi\rho}\cdot g_{\omega\gamma}
 +g_{\phi\pi\rho}\cdot g_{\phi\gamma} 
 +g_{\omega'\pi\rho}\cdot g_{\omega'\gamma}
 +g_{\omega''\pi\rho}\cdot g_{\omega''\gamma}  \nonumber \\
\label{TT} 
\eea

  This ansatz leads to  
 $\tau_{\pi^0} = 6.6(3) \cdot 10^{-17}$~s 
 for the $\pi^0\to\gamma\gamma$ lifetime
 to be compared with the PDG \cite{PDG04} value 
 $\tau_{\pi^0} = 8.3(6) \cdot 10^{-17}$~s. 
 The $\omega'$ and $\omega''$ contributions are small but not negligible and
 the value of the $\pi^0$ lifetime obtained without them reads 
$\tau_{\pi^0}(\omega \ {\rm and} \ \phi \ {\rm only})= 
 5.21(13)\cdot 10^{-17}$~s.
 Within two and a half 
 standard deviations experiment and the model are in agreement
 ($experiment-model = (1.7\pm 0.7)\cdot 10^{-17}$~s) 
 and the remaining discrepancy can be attributed to the missing
 $\rho$-resonances or a more complicated $Q^2$-dependence 
 of the propagators. One has to remember that we make here
 extrapolation from the region of $\omega,\phi,\cdots$ resonances
 where the fit was performed
 to the region of the applicability of the chiral limit.
 Thus we should recover here the chiral theory result for the
 $\pi^0\to\gamma\gamma$ lifetime (see \cite{BGP_DAPHNE})

 \bea
 \Gamma(\pi^0\to\gamma\gamma)= \frac{\alpha^2m_{\pi^0}^3}{32\pi^3f^2}\ ,
 \eea
\noindent
 which gives $\tau_{\pi^0} =8.69\cdot 10^{-17}$~s for f=132~MeV.
 Rephrasing the same statement using the coupling constants one approximately
 expects 
\bea
 \frac{1}{32\pi^4f^2} \simeq g^2_{\rho\gamma} T^2 + \cdots \ ,
\eea
where the dots correspond to terms neglected in modelling of the $3\pi$
 current. The discrepancy indicates that extending the validity of our
 model beyond the description of the $3\pi$ current one should probably take
 into account further contributions. However the 2.5 $\sigma$ discrepancy
 does not allow to draw any final conclusion. 

  The amplitude of the process $\pi^0 \to \gamma \gamma^*$ for small
  values of the four momentum of the off shell photon can be
 parameterized by a slope parameter $\alpha$
\bea
{\cal M}_{\pi^0 \to \gamma \gamma^*} = {\cal M}_{\pi^0 \to \gamma \gamma}
 \left(1 + \alpha k^{*2}\right)  \ .
\label{slope}
\eea
 Expanding the Breit-Wigner propagators one obtains 
\bea
 \alpha = \frac{1}{2}\biggl( &&\frac{1}{m_\rho^2}
 + \frac{g_{\omega\pi\rho}\cdot g_{\omega\gamma}}{T m_\omega^2}
 + \frac{g_{\phi\pi\rho}\cdot g_{\phi\gamma}}{T m_\phi^2}\nonumber\\
 && + \frac{g_{\omega'\pi\rho}\cdot g_{\omega'\gamma}}{T m_{\omega'}^2}
 + \frac{g_{\omega''\pi\rho}\cdot g_{\omega''\gamma}}{T m_{\omega''}^2}
  \biggr) \ ,
\label{slope1}
\eea
where $T$ is defined in Eq.(\ref{TT}). Numerically this
 gives $\alpha = 1.74(2)$ GeV$^{-2}$, 
 or  $\alpha\cdot m_{\pi^0}^2 = 0.0317(5)$,
 to be compared with the experimental value \cite{PDG04} 
 $\alpha\cdot m_{\pi^0}^2 = 0.032(4)$. 

 The model also predicts the decay rates for the $\rho^0\to\pi^0\gamma$,
$\omega\to\pi^0\gamma$ and $\phi \to\pi^0\gamma$: 
\bea
 \Gamma_{\rho^0\to \pi^0\gamma} =  \frac{\alpha}{24} m_{\rho}^5
  \biggl[1-\frac{m^2_{\pi^0}}{m_\rho^2}\biggr]^3 \cdot T^2
\label{rhopig}
\eea
\bea
\Gamma_{V\to \pi^0\gamma} = \frac{\alpha}{24} m_{V}^5
  \biggl[1-\frac{m^2_{\pi^0}}{m_V^2}\biggr]^3
  g^2_{V\pi\rho}\cdot g^2_{\rho\gamma} \ ,
\label{omegapig}
\eea
\noindent
 where $V$ stands for $\omega$ or $\phi$. The results are collected in 
 Table \ref{direct} and compared with the respective experimental values.

 In the $\rho^0\to \pi^0\gamma$ decay
 our 
 prediction for this branching ratio  5.2(2)$\cdot 10^{-4}$
  is in agreement with the
 6.0(1.3)$\cdot 10^{-4}$ of \cite{PDG04} within 1$\sigma$.
 As the only isospin violating effect in our model for this
 process is the charged--neutral pion mass difference,
   our prediction for the 
 ${\rm{ Br}}(\rho^{\pm}\to\pi^\pm\gamma) = 5.2(2)\cdot 10^{-4}$ is identical 
 (within the errors)  with the one for the 
neutral mode and is also in agreement with
 the data \cite{PDG04} ${\rm{ Br}}(\rho^{\pm}\to\pi^\pm\gamma) = 4.5(5)\cdot 10^{-4}$.

For the  
 $\phi\to \pi^0\gamma$ decay the
 branching ratio  ${\rm{ Br}}(\phi\to\pi^0\gamma) = 0.99(12)\cdot 10^{-3}$
 is in agreement within 2$\sigma$ with the value
 $1.23(12)\cdot 10^{-3}$ from Ref. \cite{PDG04}. Our result for 
 ${\rm{ Br}}(\omega\to\pi^0\gamma)= 15.4(5)\%$
 overestimates however the measured value (($8.92^{+0.28}_{-0.24}$)\%)
  by factor 1.7.

 The extracted couplings 
 determine also the cross section of the reaction $e^+e^-\to \pi^0\gamma$,

\bea
 &&\kern-10pt\sigma(e^+e^- \to \pi^0 \gamma) =  \frac{2\pi^2\alpha^3}{3}
 \left(1-\frac{m_{\pi^0}^2}{s}\right)^3g^2_{\rho\gamma}\nonumber \\
 \cdot && |\
 \bigl[ 
 g_{\omega\pi\rho}\cdot g_{\omega\gamma}
+g_{\phi\pi\rho}\cdot g_{\phi\gamma} \nonumber \\
 &&+g_{\omega'\pi\rho}\cdot g_{\omega'\gamma}
 +g_{\omega''\pi\rho}\cdot g_{\omega''\gamma}  
 \bigr] \cdot  BW_{\rho}(s) \nonumber\\
&&+g_{\omega\gamma}\cdot
 g_{\omega\pi\rho} 
 \cdot  BW_{\omega}(s)
 +g_{\phi\gamma}\cdot
 g_{\phi\pi\rho}
 \cdot  BW_{\phi}(s) \nonumber \\
 &&+g_{\omega'\pi\rho}\cdot g_{\omega'\gamma}
 \cdot BW_{\omega'}(s)
  +g_{\omega''\pi\rho}\cdot g_{\omega''\gamma}
 \cdot BW_{\omega''}(s) 
  |^2 \ ,\nonumber \\
\label{sigeepi0g}
\eea
 with $BW_i \  , i =\rho,\omega,\phi,\omega',\omega'' $
 defined in Eq.(\ref{BW}).

\begin{figure}[ht]
\begin{center}
\epsfig{file=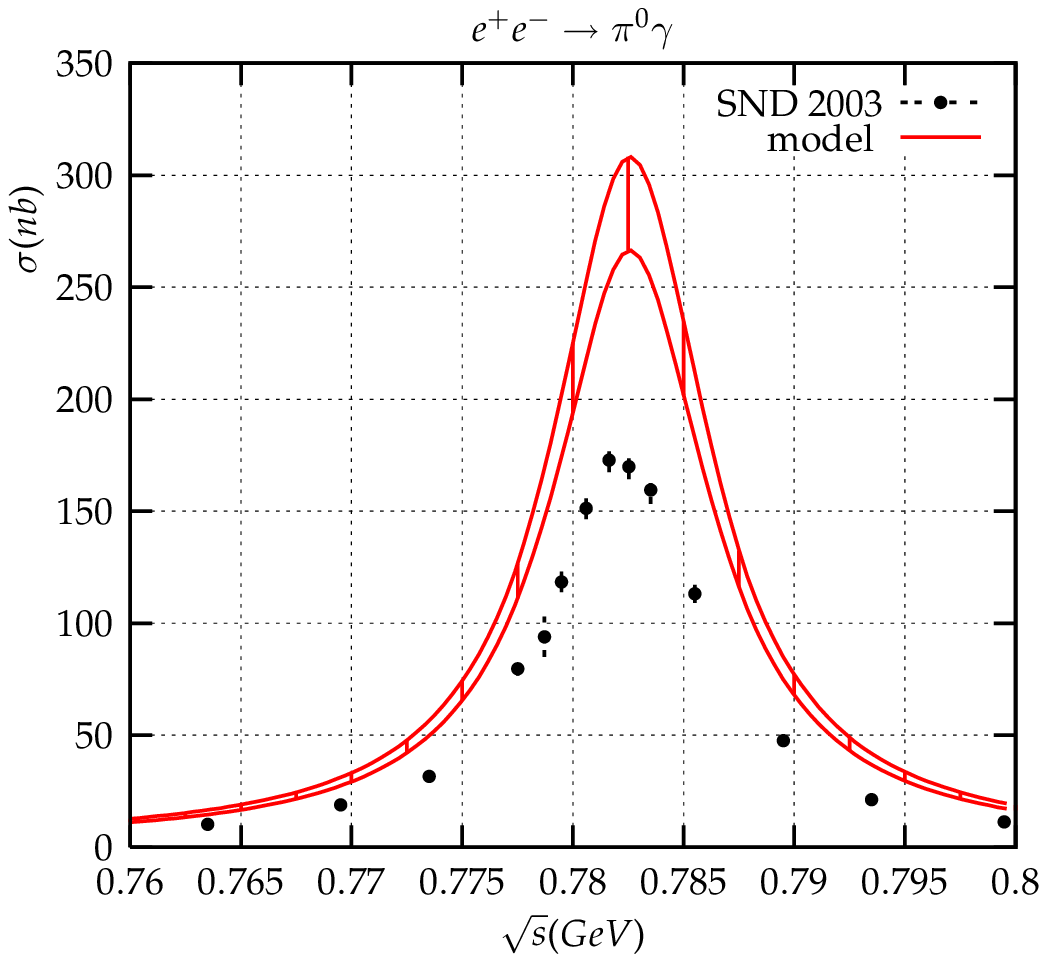,width=7.5cm}
\epsfig{file=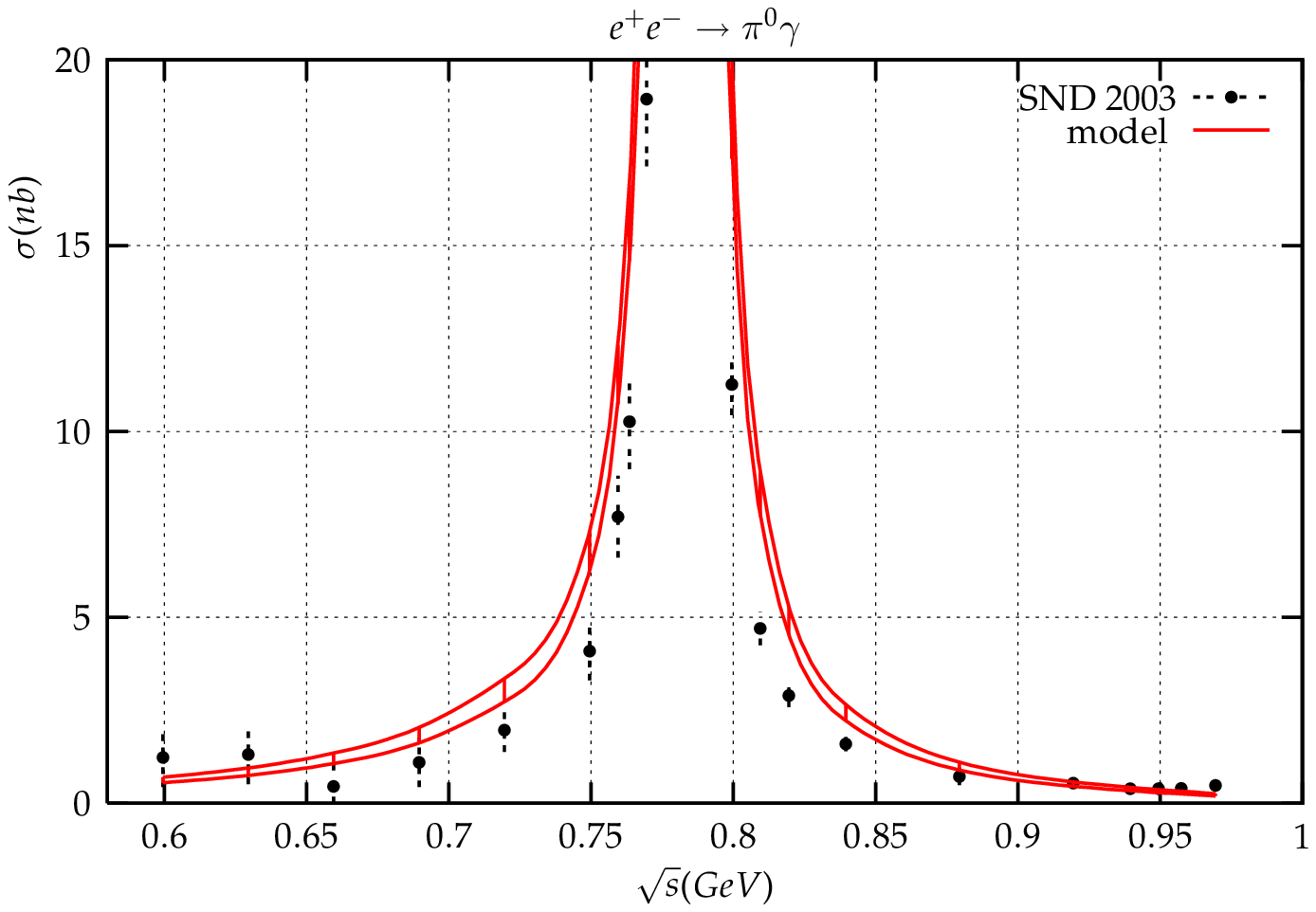,width=7.5cm}
\epsfig{file=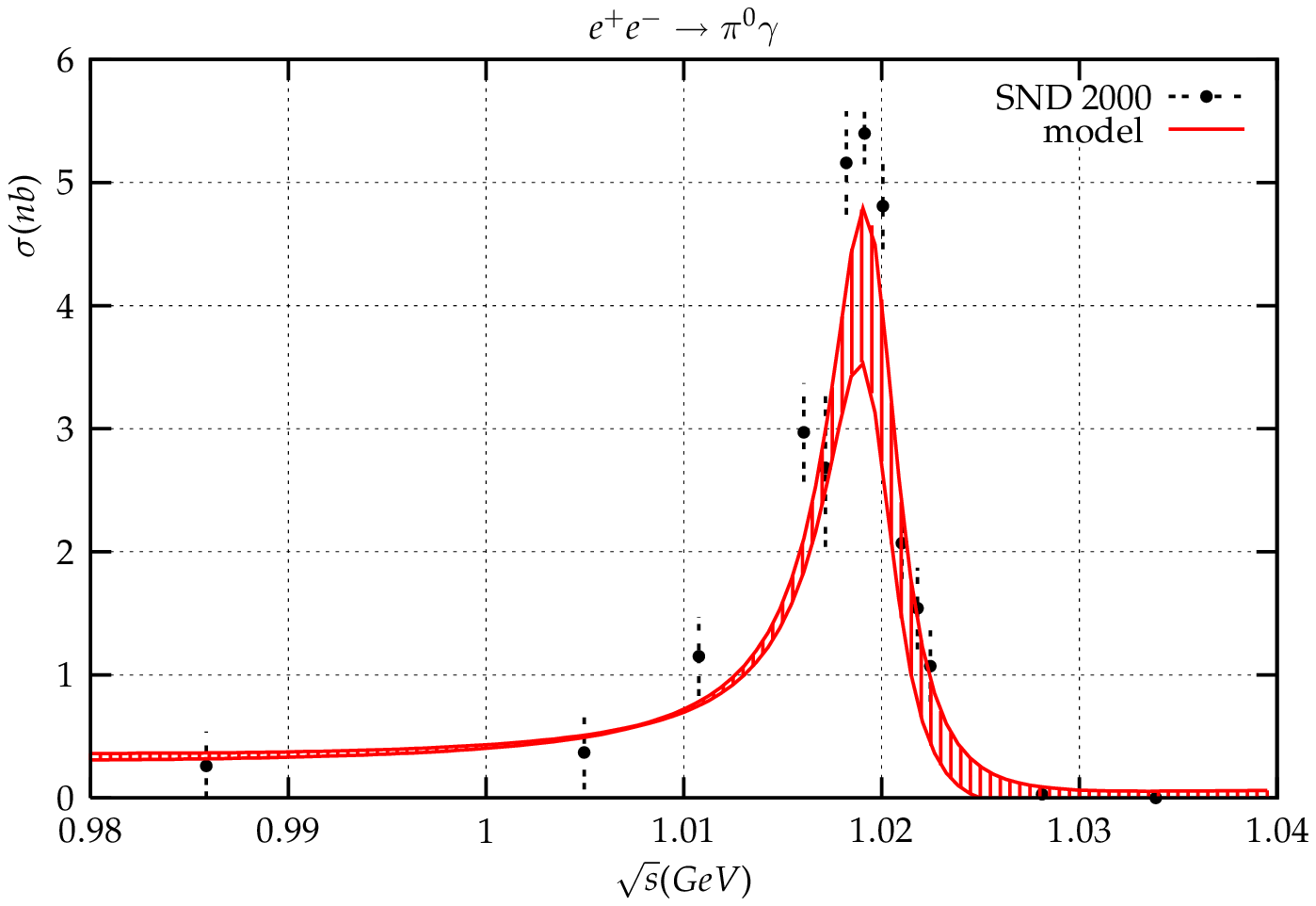,width=7.5cm}
\caption{Differential cross section of the process 
$e^+ e^- \to \pi^0\gamma $. Data from Refs.\cite{Achasov:03pig} (SND 2003) and
\cite{Achasov:00} (SND 2000) are shown together with 1 $\sigma$ allowed
 bands for our model predictions.
}
\label{cspigam}
\end{center}
\end{figure}

 The comparison with existing data \cite{Achasov:03pig,Achasov:00}
 is shown in Fig. \ref{cspigam}, where
 one standard
 deviation bands are given.
 Reasonable agreement is observed around the $\phi$ resonance
 in contrast to the $\omega$ region. This is of course a reflection
 of the agreement and disagreement of the corresponding decay rates.
 A similar behaviour was also observed
 in Ref. \cite{RFPP}. 
Their
model predict a value for  $\Gamma(\omega\to\pi^0\gamma)$ well in accordance
with the experimental value, but could not easily reproduce
$\Gamma(\omega\to \pi^+\pi^-\pi^0)$.
 Further, both theoretical
 and experimental, studies of reactions $e^+e^- \to \pi^+\pi^-\pi^0$
 and $e^+e^- \to \pi^0\gamma$ are thus required.

The branching ratios of the $\omega$, $\phi$ and $\rho$ decays to
 $\pi^+\pi^-\pi^0$ obtained within our model
 can be found in Table \ref{br3pi}. They  are in agreement with 
  the PDG values \cite{PDG04} for $\omega$, $\phi$
 within 1-2 standard deviations. However the predicted
 value for ${\rm{ Br}}(\rho^0\to\pi^+\pi^-\pi^0)$ is more then two orders 
 of magnitude smaller than the experimental value 
 ${\rm{ Br}}(\rho^0\to\pi^+\pi^-\pi^0) \simeq 1\cdot 10^{-4}$, which is consistent with
 zero at two sigma level.

\begin{table}[hb]
\begin{center}
\caption{Branching ratios of the $\omega$, $\phi$ and $\rho$ to
 $\pi^+\pi^-\pi^0$ decays: model vs. experiment \cite{PDG04}.}
\label{br3pi}
\begin{tabular}{l|l|l}
 & model& experiment \\ \hline
 ${\rm{ Br}}(\omega\to\pi^+\pi^-\pi^0)$ &  95.1(22)\%& 89.1(7)\%
  \\ \hline
 ${\rm{ Br}}(\phi\to\pi^+\pi^-\pi^0)$ &  14.5(22)\%& 15.4(5)\%\\ \hline
 ${\rm{ Br}}(\rho\to\pi^+\pi^-\pi^0)$ &  1.9(3)$\cdot$ 10$^{-6}$&
 ($1.01^{+0.54}_{-0.36}$$\pm$0.34)$\cdot$ 10$^{-4}$\\ \hline
 \end{tabular}
\end{center}
\end{table}

\section{Tests of the MC code }

 In comparison to the previous versions of PHOKHARA
 the default random number generator was changed to the double precision
 version of RANLUX \cite{RANLUX} written in C by Martin L\"uscher. A 
  C--FORTRAN interface is provided with the PHOKHARA 5.0 distribution.

 To assure a technical precision of the
 code better than a fraction of one per mile
 a number of tests were performed for the new hadronic state
 in the generator.  Among other tests, the initial state emission
 was tested against known analytical results, where all photon angles
 are integrated, similarly to previously performed tests for other
 hadronic channels (see \cite{Rodrigo:2001kf,Czyz:2002np,Czyz:PH03,Czyz:PH04}).
 The independence of the result from the soft photon separation parameter
 was also checked.


\section{Summary and Conclusions}

 The Monte Carlo event generator PHOKHARA
 has been extended  to the three-pion mode. The model 
 adopted for the hadronic form factor properly describes the currently
 available data for the cross section and the distributions. The ansatz
 is based on  generalized vector dominance and is consistent with various other
 measurements like $\Gamma(\pi^0\to\gamma\gamma)$, the slope parameter
 of the $\pi^0\to\gamma\gamma^*$ amplitude and radiative vector meson
 decays $\rho\to \pi^0\gamma$, $\phi\to \pi^0\gamma$,
 but is in conflict with  $\omega\to \pi^0\gamma$.

 The current version of the computer program (PHO\-KHA\-RA 5.0) is
 available at

  {\tt http://cern.ch/german.rodrigo/phokhara}.


\section*{Acknowledgements}
The authors thank J. Portol\'es for very helpful discussions
and comments on the manuscript.


\end{document}